\documentclass[sigconf]{acmart}
\usepackage{multirow}
\usepackage{caption} 
\usepackage{graphicx}
\usepackage{subcaption}
\AtBeginDocument{%
  }

\setcopyright{cc}
\copyrightyear{2026}
\acmYear{2026}
\acmDOI{XXXXXXX.XXXXXXX}
\acmISBN{978-1-4503-XXXX-X/2018/06}

\begin{document}

\title{Task-Aware Automated User Profile Generation for Recommendation Simulation Using Large Language Models}


\author{Xinye Wanyan}
\email{s4142700@student.rmit.edu.au}
\orcid{0009-0002-7264-1803}
\affiliation{%
  \institution{RMIT University}
  \city{Melbourne}
  \state{VIC}
  \country{Australia}
}
\author{Chenglong Ma}
\email{chenglong.ma@rmit.edu.au}
\orcid{0000-0002-6745-4029}
\affiliation{%
  \institution{RMIT University}
  \city{Melbourne}
  \state{VIC}
  \country{Australia}
}
\author{Danula Hettiachchi}
\email{danula.hettiachchi@rmit.edu.au}
\orcid{0000-0003-3875-5727}
\affiliation{%
  \institution{RMIT University}
  \city{Melbourne}
  \state{VIC}
  \country{Australia}
}
\author{Ziqi Xu}
\email{ziqi.xu@rmit.edu.au}
\orcid{0000-0003-1748-5801}
\affiliation{%
  \institution{RMIT University}
  \city{Melbourne}
  \state{VIC}
  \country{Australia}
}

\author{Jeffrey Chan}
\email{jeffrey.chan@rmit.edu.au}
\orcid{0000-0002-7865-072X}
\affiliation{%
  \institution{RMIT University}
  \city{Melbourne}
  \state{VIC}
  \country{Australia}
}

\begin{abstract}
Large Language Model (LLM)-based agent simulation has emerged as a promising approach to meet the increasing demand for real-time and rigorous evaluation in modern recommender systems. A typical LLM-driven simulation framework comprises three essential components: the profile module, memory module, and action module. However, existing studies have primarily concentrated on enhancing the memory and action modules, with limited attention to profile generation, which plays a pivotal role in ensuring realistic agent behaviours and aligning simulated interactions with real user dynamics. Moreover, the scarcity of datasets specifically designed for recommendation simulations has led to heavy reliance on manually crafted profiles, significantly limiting the scalability and generalisability of simulation frameworks across different datasets. To address these challenges, this work proposes an \underline{A}utomated \underline{P}rofile \underline{G}eneration Framework for \underline{Rec}ommendation \underline{Sim}ulation, APG4RecSim, that constructs realistic, coherent, and robust user profiles with minimal supervision. 
Extensive experiments on three benchmark datasets demonstrate that APG4RecSim achieves the best overall performance on discrimination, ranking, and rating tasks, improving ranking quality by up to 7\% in nDCG@10 and reducing rating distribution divergence by 8\% in JSD compared to existing profile-generation baselines.
Beyond overall performance gains, our results show that profiles generated by APG4RecSim are resilient to popularity- and position-induced biases and maintain stable performance across datasets and different LLMs.
\footnote{The source code is available at~\url{https://github.com/WennyXY/APG4Sim}.}
\end{abstract}




\maketitle

\section{Introduction}
Modern recommender systems (RSs) require real-time user interactions to facilitate effective performance evaluation and continuous system refinement \cite{gao2022kuairec,deldjoo2024review,zhang2020evaluating}. To address the limitations of static datasets and costly user studies, LLM-based agent simulation has emerged as a promising approach for generating instantaneous, realistic user feedback and supporting large-scale, concurrent evaluation environments for RSs \cite{zhao2024recommender,zhang2025survey,RecUserSim2025Chen,beyond2025jin,evalrec25xu}. A typical LLM-based agent consists of three fundamental components, with the LLM serving as the core reasoning engine \cite{chen2019generative, wang2025recagent,RecUserSim2025Chen,beyond2025jin}. The \textbf{profile module} encodes the agent's identity, preferences, and behavioural traits, thereby defining how the agent tends to act \cite{hu2024quantifying,Profiling2025Wang}. The \textbf{memory module} maintains the agent's interaction history, allowing it to reason with accumulated context rather than behaving independently at each time step \cite{llmmemory2025Zhang}. The \textbf{action module} takes the recommendations as input and generates decisions conditioned on the profile and memory, ultimately determining how the agent interacts with the RS throughout the simulation loop.

Despite agent behaviour being governed by the synergy of three distinct modules, existing research primarily focuses on exploring the design of the memory and action modules \cite{wang2025recagent,agent4rec,Zhang_2025_LLMpowerSim4Rec}, while largely overlooking profile generation. RecAgent \cite{wang2025recagent} advances user simulation by implementing a hierarchical memory architecture designed to mirror human cognitive processes. Agent4Rec \cite{agent4rec} proposes a hybrid memory system that intertwines objective interaction history with subjective emotional memories, allowing the agent to mimic the nuance of human decision-making. 
\citet{Zhang_2025_LLMpowerSim4Rec} enhances the interaction inference of the action module by combining an LLM-based logical model with two statistical models to ensure the robustness and reliability of agent behaviours.
However, current frameworks typically approach profile construction by manually engineering a set of target attributes before employing LLMs for extraction and inference. 
Empirical evidence demonstrates that guided profile generation significantly enhances personalisation across diverse tasks \cite{zhang2024guided}, highlighting its crucial role in advancing the reliability and effectiveness of user behaviour simulation within RSs \cite{Bhattacharjee2024}. In LLM-based simulation, these findings highlight the limitations of unguided or purely heuristic profile extraction and motivate the need for structured, context-aware approaches that align profiles with downstream decision-making requirements.

The oversight of profile construction and optimisation introduces three fundamental \textbf{limitations} to current simulation frameworks: 
(1) \textbf{Information Bottleneck due to Rigid Schema}: By relying on manually defined attribute lists, existing methods suffer from a ``closed-world'' constraint. This prevents the agent from capturing latent or evolving user traits (e.g., shifting interests or complex conditional preferences) that fall outside the designer's static taxonomy. 
(2) \textbf{Behavioural Misalignment}: Without a rigorous, statistically grounded profiling mechanism, agents often lack a strong ``instructional anchor''. Poorly constructed profiles directly degrade the realism of simulated behaviours and weaken the alignment between simulated and real user interactions. 
(3) \textbf{Lack of Domain Generalisability}: Manual profile engineering is highly domain-dependent. A framework optimised for e-commerce attributes (e.g., price, brand) fails to generalise to content streaming (e.g., genre, mood) without significant human re-engineering, limiting the scalability of the simulation.
Existing profiling methods fail to account for the situational relevance of user traits; an attribute critical for one recommendation task (e.g., price sensitivity in e-commerce) may be irrelevant noise in another (e.g., content rating in social media). Furthermore, the granularity of user traits must align with the specific metadata available in the target domain to be actionable.



Based on the identified limitations of existing studies, this work investigates the following research questions:
\begin{itemize}
    \item \textbf{RQ1 } How to automatically construct user profiles from interaction data without manually defined attribute schemas, while capturing latent and task-relevant user traits through context-aware consolidation?
    \item \textbf{RQ2} How do automatically generated profiles improve the behavioural alignment between simulated agents and real users across multiple recommendation tasks?
    \item \textbf{RQ3 } Can the proposed profile generation framework generalise across domains, datasets, and LLM backbones under a unified and controlled evaluation protocol?
\end{itemize}
We answer these questions through the design of our framework and extensive experimental evaluation.
In summary, the main contributions of this paper can be concluded as follows:
\begin{itemize}
    \item We propose an automated profile generation framework that constructs diverse and coherent user profiles from interaction data without relying on manually engineered attribute schemas.
    \item We introduce a context-aware consolidation mechanism that aligns profile attributes with task-specific decision processes, improving behavioural realism in agent-based simulation. 
    \item We design a systematic evaluation protocol that isolates profile quality from agent architectures and demonstrate consistent improvements compared to existing profile generation methods across ranking, discrimination, and rating tasks on multiple datasets and LLM backbones.
\end{itemize}

\section{Related Work}
\subsection{Agent-based Simulation for RSs}
Generating synthetic interaction data through agent-based simulation offers a practical solution to several challenges in RSs.
By leveraging the advanced reasoning and language understanding capabilities of LLMs \cite{radford2019language}, these approaches enable agents to perform complex and realistic behaviours interacting with online systems \cite{Park2023Simulacra,wang2025recagent,agent4rec,sigir2025Zhang}.
\citet{sigir2025Ungruh} critique the ``monolith'' approach which relied on monolithic behavioural assumptions that treated users as uniform decision-makers, demonstrating that incorporating diverse behavioural choice models, a ``mosaic'' of user patterns, is essential for high-fidelity simulation.
RecAgent \cite{wang2025recagent} primarily focuses on memory module and action module designing a three-level hierarchical memory module combined with recommendation-related behaviour and two types of social behaviour.
Agent4Rec \cite{agent4rec} proposes to integrate both factual and emotional memories and an emotion-driven reflection mechanism for decision-making which supports both taste-driven and emotion-driven actions in RSs.
\citet{Zhang_2025_LLMpowerSim4Rec} combine an LLM-based logical model with statistical models for binary interactions of \textit{like} or \textit{dislike} inference.
SimUSER \cite{bougie2025simuser} leverages LLM-based agents to simulate user behaviour and analyse the impact of thumbnails, exposure effects, and reviews on user engagement within a standard modular architecture comprising profiling, memory, planning, and action components.
\citet{sigir2025Cai} propose the Agentic Feedback Loop (AFL) framework, which models a collaborative dialogue between the system and the user to mutually enhance performance. Notably, their user agent bypasses explicit natural language profiling, instead relying on a pre-trained reward model (e.g., SASRec) to dictate preferences, effectively limiting the LLM's role to the narration of predetermined scores.
DyTA4Rec \cite{DyTa4Rec} focuses on incorporating temporal reasoning to infer users’ dynamic behaviour in recommender systems.

Existing agent-based simulation studies predominantly focus on improving agent reasoning, memory, or planning mechanisms, while user profile construction is often assumed to be manual, static, and weakly evaluated.

\begin{figure*}
    \centering
    \includegraphics[width=1\linewidth]{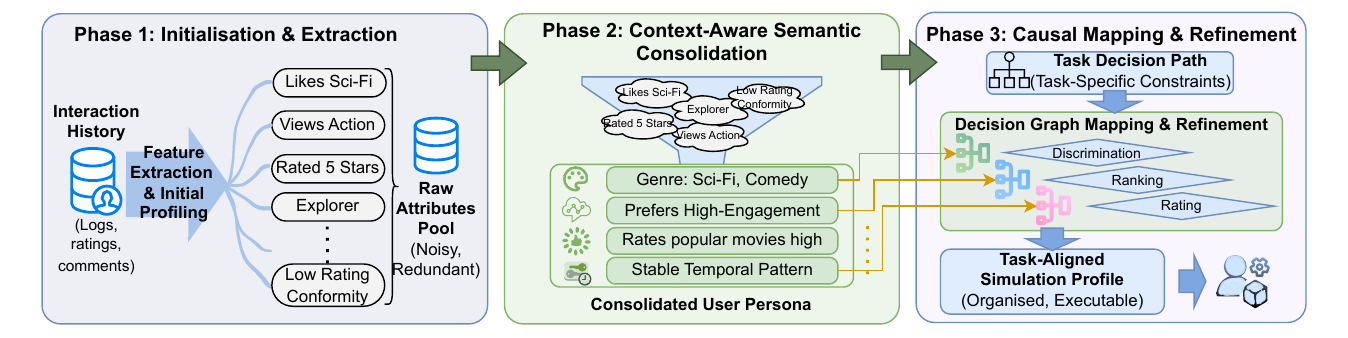}
    \caption{Overview of APG4RecSim, a training-free and context-adaptive LLM-based profile generation workflow for recommender system simulation. This diagram illustrates the proposed three-stage methodology for creating high-fidelity profiles for agents. Stage 1 (Initialisation \& Extraction) involves mining raw features from a user's interaction history (e.g., logs, ratings, reviews) to create a pool of attributes which may contain noises. Stage 2 (Context-Aware Semantic Consolidation) employs a semantic merging and deduplication process to refine the raw attributes into a concise, coherent user profile based on given context. Finally, Stage 3 (Decision Path Mapping \& Refinement) maps these consolidated attributes to specific nodes in a decision path, resulting in a task-aligned simulation profile with executable policies for a simulation agent.}
    \label{fig:placeholder}
\end{figure*}
\subsection{User Profile in RS Simulation}
User profiling is a foundational component of recommender systems, traditionally represented through feature vectors, latent embeddings, or demographic attributes derived from historical interactions \cite{ontological2004,demographic2016recsys,lettingo2025Wang}. These representations are primarily designed to optimise recommendation accuracy rather than to support behavioural simulation, and they are often opaque or difficult to interpret in simulation settings.
RecAgent \cite{wang2025recagent} designs user profile with attributes consisting of personality, interests, and behaviour features. They mention three strategies to generate user profiles for different needs: handcraft, GPT-based, and real-data alignment methods but don't delve deep into it.
\citet{ma2025PUB} propose a personality-driven user behaviour simulator powered by LLMs integrating the Big Five personality traits to model personalised behavioural patterns in recommender systems.
However, these studies have not delved deep into the profile generation. They only provide the attributes they design, and haven't explored the evaluation, reason, and optimisation. 
\citet{Zhang_2025_LLMpowerSim4Rec} consider the importance of the agent profiling and propose to utilise LLM as the evaluator to assess the alignment between profile and interaction history.
\citet{UserIPTuning2025}
introduce UserIP-Tuning to optimise user profiling via continuous latent vectors, effectively bypassing the token constraints and noise inherent in textual history. However, the learned soft prompts are opaque embeddings that cannot be directly utilised as natural language profiles for standard LLM agents.

Despite the central role of profiles in shaping agent behaviour, user profiles are rarely treated as a first-class research object in recommender system simulation.
Even recent LLM-based profile generation approaches typically rely on predefined attribute schemas, limiting their ability to adapt across tasks and domains and constraining behavioural generalisation. In contrast, our work centres on automated and context-aware profile generation, introducing a training-free and generalisable framework that explicitly aligns user profiles with task-specific decision processes to enable more realistic user behaviour simulation.

Leveraging these insights, we introduce a multi-stage, context-sensitive framework for guided profile generation. This approach automates the construction of user profiles without requiring additional training, ensuring explicit alignment with task-specific decision logic to facilitate more consistent, realistic, and generalisable behaviour simulation.
\section{Methodology}

We propose a three-stage framework, APG4RecSim, to convert raw user interaction histories into \emph{task-executable} agent profiles for recommendation simulation.
Given a user history $\mathcal{H}_u$, our goal is to produce (i) a compact, domain-grounded description of stable user traits, and (ii) task-specific decision policies that operationalise these traits for a target simulation task (e.g., discrimination, ranking, or rating).
Accordingly, the pipeline consists of: (1) \textbf{Attribute Initialisation and Extraction}, which elicits a high-recall set of candidate traits from $\mathcal{H}_u$; (2) \textbf{Context-Aware Semantic Consolidation}, which deduplicates and grounds traits to the dataset schema under an explicit context $\mathcal{C}$; and (3) \textbf{Causal Mapping and Refinement}, which instantiates a task decision structure and maps consolidated traits to executable policies via a counterfactual relevance test.


\subsection{Problem Definition}

For each user $u$, we define the interaction history $\mathcal{H}_u = \{(i_k, r_k, t_k)\}_{k=1}^{n}$, where each tuple consists of an item $i_k$, the associated feedback signal $r_k$ (e.g., rating, click, or review), and a timestamp $t_k$.
We define a \emph{Task-Aligned User Profile} as a tuple
\[
\mathcal{P}_u(\mathcal{T}) = \langle \mathcal{I}_{meta}, \mathcal{A}_{dom}, \mathcal{G}_{policy}(\mathcal{T}) \rangle,
\]
where $\mathcal{I}_{meta}$ stores immutable identifiers (e.g., user ID, optional demographic tags), $\mathcal{A}_{dom}$ is a domain-grounded set of consolidated natural-language traits (task-agnostic), and $\mathcal{G}_{policy}(\mathcal{T})$ is a set of task-specific decision policies parameterised by task specification $\mathcal{T}$.
Our objective is to learn a mapping
\[
f:\ \mathcal{H}_u \rightarrow \mathcal{P}_u(\mathcal{T}),
\]
implemented as a staged decomposition:
\[
\begin{aligned}
&\mathcal{A}_{raw} \leftarrow g_{\text{ext}}(\mathcal{H}_u);\quad
\mathcal{A}_{dom} \leftarrow g_{\text{merge}}(\mathcal{A}_{raw},\mathcal{C});\\
&\mathcal{G}_{policy}(\mathcal{T}) \leftarrow g_{\text{map}}(\mathcal{A}_{dom},\mathcal{T}).
\end{aligned}
\]

\subsection{Stage 1: Initialisation and Extraction}
The objective of the first stage is to construct a high-recall candidate pool of user traits. To mitigate the stochastic nature of LLM generation and ensure comprehensive coverage of latent preferences, we do not rely on a single generation pass. Instead, we initialise $N=3$ distinct preliminary profiles in parallel. This ensemble size is empirically motivated; our preliminary experiments indicate that trait recall saturates at 2–3 initialisations, making $N=3$ sufficient to capture the user's persona without incurring unnecessary computational overhead.
We prompt an LLM to explain the observed history $\mathcal{H}_u$ by generating natural-language descriptors that account for recurring interaction patterns, rather than predicting future items.
Formally, we aggregate these into a raw feature pool:
$$\mathcal{A}_{raw} = \bigcup_{j=1}^{3} \mathcal{A}^{(j)}$$
\[
\mathcal{A}^{(j)} = g_{\text{ext}}(\mathcal{H}_u;\,\theta) = \text{LLM}(\mathcal{H}_u \mid \text{Prompt}_{\text{ext}}),
\]
where $\mathcal{A}_{raw}$ is intentionally over-complete and may contain redundancy, paraphrases, and attributes not aligned with the dataset vocabulary.
This union strategy prioritises recall to minimise missed latent preferences, accepting that the pool will contain redundancy and noise, which are subsequently resolved in the domain-aware Stage~2.

\subsection{Stage 2: Context-Aware Consolidation}
Raw attributes $\mathcal{A}_{raw}$ often exhibit semantic redundancy (e.g., paraphrases) and vocabulary mismatch with the target dataset.
We consolidate $\mathcal{A}_{raw}$ into a compact, domain-grounded trait set $\mathcal{A}_{dom}$ by conditioning the merging process on an explicit context
\[
\mathcal{C} = \langle \mathcal{D}_{info}, \mathcal{T}_{desc}, \mathcal{E}_{shot} \rangle,
\]
where $\mathcal{D}_{info}$ describes the dataset schema and admissible item metadata fields, $\mathcal{T}_{desc}$ specifies the simulation task family and the required semantic granularity, and $\mathcal{E}_{shot}$ provides a representative example instance to calibrate terminology.
We then apply LLM-guided semantic consolidation:
\[
\mathcal{A}_{dom} = g_{\text{merge}}(\mathcal{A}_{raw},\mathcal{C}) = \text{LLM}(\mathcal{A}_{raw} \mid \mathcal{C}, \text{Prompt}_{\text{merge}}),
\]
which performs (i) deduplication into canonical descriptors and (ii) grounding to dataset-native concepts, producing a task-agnostic profile ready for policy mapping.

\subsection{Stage 3: Causal Mapping and Refinement}
To make traits executable for simulation, we instantiate a task decision path and map each trait to the decision step(s) it directly affects.

\subsubsection{Decision Path Instantiation}
For a task $\mathcal{T}$ (discrimination, ranking, or rating), we represent the agent's decision structure as an ordered \emph{decision path} $\Pi(\mathcal{T}) = [s_1,\dots,s_m]$, where each step $s_j$ corresponds to a cognitively meaningful judgement (e.g., hard filtering, trade-off comparison, final choice/rating).
The path is obtained either from a domain-expert heuristic template or generated by an LLM conditioned on the task description and a single input-output exemplar.

\subsubsection{Counterfactual Trait-to-Step Mapping}
We map a trait $a \in \mathcal{A}_{dom}$ to a step $s_j$ if altering only $a$ (while keeping all other traits and the candidate set fixed) changes the agent's output at that step under the task interface.
Operationally, we create a counterfactual profile $\mathcal{A}_{dom}^{(a \leftarrow a')}$ by replacing $a$ with a minimally perturbed variant $a'$ (e.g., negation or weakened preference), run the agent on the same instance, and mark $a$ as relevant to $s_j$ if the predicted decision differs (selection set changes for discrimination, ranking order changes for ranking, or rating changes beyond a small threshold for rating).
Traits that never trigger a counterfactual change are deprioritised or merged further, whereas traits with verified counterfactual influence are preserved to avoid over-pruning.
The final output is a task-aligned policy set $\mathcal{G}_{policy}(\mathcal{T})$ that attaches each retained trait to the step(s) it affects, enabling controllable and interpretable simulation.

\section{Experiments}
This section presents a comprehensive experimental evaluation of the proposed framework, detailing the experimental setup, baseline comparisons, and results across multiple tasks and datasets. Furthermore, we conduct rigorous ablation studies, robustness analyses, and an investigation into the impact of LLM parametric priors on simulation.

\subsection{Experimental Settings}
\subsubsection{Datasets and Data Split} 
We conduct experiments on three public recommendation datasets, including MovieLens-1M (ML-1M) \cite{movielens_1m}, Amazon-Book \cite{hou2024bridging}, and Amazon-Beauty \cite{hou2024bridging}.
For each dataset, interactions are ordered chronologically by timestamp. For each user, the earliest 80\% of interactions are used for profile generation, while the remaining 20\% are reserved for profile evaluation.

\subsubsection{Baselines and Evaluation Settings} 
We compare our proposed approach with three established profile-generation baselines across three tasks and three benchmark datasets. 
(1) \textbf{Recent-Interaction Baseline}: the most recent raw interacted items are directly used as contextual input for the agent to perform task prediction.
(2) \textbf{RecAgent} \cite{wang2025recagent}: an LLM-based approach that extracts a fixed set of pre-defined user attributes from the interaction history, consisting of personality, interests, and behaviour features. 
This approach categorises users into five distinct behavioural types, including watcher, explorer, critic, chatter, or poster, and prompts the LLM to classify each user based on the predefined roles and the user's historical interactions.
(3) \textbf{Agent4Rec} \cite{agent4rec}: a hybrid framework combining statistical metrics with LLM-derived traits, we adapt this baseline to focus strictly on its generative capabilities. In our implementation, only the LLM-extracted features, specifically the user tastes and their associated rationales are retained to ensure a fair, content-centric comparison with our proposed method.

To evaluate the performance of profiles, we create three simulation scenarios for different interaction tasks, including interacted items discrimination, rating, and ranking tasks.

The \textbf{discrimination task} is to assess the agent's ability to distinguish interacted items from non-interacted ones under controlled candidate sets. 
For each test instance, we construct a candidate pool of size $C=10$, containing $P = 1, 3, 5$ ground-truth positive items randomly drawn from the test set to simulate varying preference patterns. The remaining $C-P$ items, drawn from non-interacted items, are used as distractors. 
Given a recommendation list, the agent's task is to use the learned profile to distinguish all items that belong to the ground-truth positive set. The task evaluates whether the agent can correctly identify and prioritise the ground-truth items over non-relevant ones. We use the Overlap Ratio to measure the proportion of ground-truth items that the agent can discriminate.

The \textbf{ranking task} evaluates the agent's ability to identify the most relevant item given a user profile. Specifically, we adopt a top-1 ranking setting following previous work \cite{hou2022UniSRec,zhang2024agentcf,zhou2020s3,DyTa4Rec}, where a single positive item is sampled from the evaluation set and mixed with a set of negative items. Conditioned on the provided user profile, the agent is required to rank all candidate items and assign the highest rank to the positive item. 
Its ranking performance is measured by the position of the held-out target item within this ranking. Standard ranking metrics, including nDCG@5, nDCG@10, and Hit Rate@3 are computed to quantify how effectively the agent identifies the user's real next choice among the candidate items.

The \textbf{rating task} evaluates the agent's ability to predict users' explicit feedback based on the given profile. For each test instance, the agent is provided 10 target items sampled from the evaluation set and is asked to generate ratings conditioned on the user profile and item information. The predicted rating is then compared against the ground-truth user rating. To comprehensively assess rating performance, we evaluate results at both the micro level and the macro level.
At the micro level, we adopt Root Mean Squared Error (RMSE) to measure the deviation between the rating predicted by the agent and the corresponding ground-truth rating for the same item, thereby quantifying the accuracy of individual rating predictions.
At the macro level, we assess the overall alignment between simulated and real rating behaviours by grouping items according to their rating values. For each rating group, we compute the Jensen–Shannon Divergence (JSD) \cite{jsd1991} to quantify the discrepancy between the rating distributions produced by the agent and the ground-truth distributions. This macro-level evaluation captures not only numerical accuracy but also the consistency of global rating patterns, providing a complementary perspective to the micro-level analysis.

\subsubsection{Implementation Details}
In our experiments, we primarily employ gpt-4o-mini (temperature is set to 0.1) as the reasoning agent to simulate user behaviour conditioned on the generated profiles. 
We also conduct generalisation experiments using Llama-3.3-70B-Instruct, GPT76-5.1, and DeepSeek-V3.2 to further assess the framework's robustness across different LLM architectures.
As our study focuses on evaluating the quality and effectiveness of user profiles rather than agent architecture design, the agent is not equipped with additional components (e.g., memory, action, or planning modules) beyond profile-conditioned reasoning. The LLM is prompted with the user profile and instructed to role-play the corresponding user when interacting with recommended items. 
To mitigate the risk of propagating stereotypical biases often observed in recommendation algorithms \cite{chakraborty2017makes,biasRecSys2023,genderbias2025}, we enforce a strictly behavioural profiling strategy. We deliberately exclude explicit demographic attributes (e.g., gender, age, location) from both the profile generation and behavioural inference phases, ensuring the agent's decisions are grounded solely in observed interaction patterns rather than demographic priors.
To ensure consistency across experiments, we employ a fixed context window for profile generation. Specifically, all user profiles are initialised based on the most recent items from the interaction history, except where noted otherwise.
All experiments are conducted over five independent runs, and results are reported as averages to ensure robustness and reduce variance.

\subsection{Main Results}

\begin{table*}
    \centering
    \caption{Discrimination ability comparison of different profile-generation methods across three benchmark datasets. Performance is evaluated using Overlap Ratio under three discrimination settings, including discriminate 1,3, and 5 positive item(s) from 10 recommended items, where the agent is required to identify 1, 3, or 5 ground-truth items from a candidate list of 10 items. Best results are bolded; second-best are underlined. }
    \label{tab:discrimination}
    \begin{tabular}{c|ccc|ccc|ccc}
        \toprule
        \multirow{2}{*}{Model} 
        & \multicolumn{3}{c}{ML-1M} & \multicolumn{3}{|c}{Amazon-Book} & \multicolumn{3}{|c}{Amazon-Beauty} \\
        & 1:10 & 3:10 & 5:10 
        & 1:10 & 3:10 & 5:10
        & 1:10 & 3:10 & 5:10 \\
        \midrule
        No Profile  
        & 0.1100 & 0.2495 & 0.5650
        & 0.1500 & 0.3415 & 0.5750
        & 0.0700 & 0.2490 & 0.4620 \\
        RecAgent
        & 0.0900 & 0.2745 & 0.5100
        & 0.1700 & 0.5930 & 0.7300
        & 0.1000 & 0.2560 & 0.4360 \\
        Agent4Rec
        & 0.1000 & 0.2750 & 0.5200
        & 0.4250 & 0.5925 & \underline{0.7380}
        & 0.1500 & \underline{0.4665} & \textbf{0.5700} \\
        \midrule
        Semantic Merge 
        & \underline{0.1900} & \underline{0.3760} & \textbf{0.5740}
        & \underline{0.4300} & \underline{0.6175} & 0.7280
        & \underline{0.2900} & 0.4470 & 0.5120 \\
        
        APG4RecSim
        & \textbf{0.2000} & \textbf{0.3955} & \underline{0.5665}
        & \textbf{0.4500} & \textbf{0.6270} & \textbf{0.7400}
        & \textbf{0.3300} & \textbf{0.4710} & \underline{0.5380} \\
        \bottomrule
    \end{tabular}
    
\end{table*}

\begin{table*}
    \centering
    \caption{Ranking performance comparison on ML-1M, Amazon-Books, and Amazon-Beauty, where agents are conditioned on user profiles generated by different profile-generation models. Performance is evaluated using nDCG@5, nDCG@10, and Hit Rate@3 (HR@3). Best results are bolded; second-best are underlined.}
    \label{tab:ranking}
    \begin{tabular}{c|ccc|ccc|ccc}
        \toprule
        \multirow{2}{*}{Model} 
        & \multicolumn{3}{c}{ML-1M} & \multicolumn{3}{|c}{Amazon-Book} & \multicolumn{3}{|c}{Amazon-Beauty} \\
        & nDCG@5 & nDCG@10 & HR@3 & nDCG@5 & nDCG@10 & HR@3 & nDCG@5 & nDCG@10 & HR@3  \\
        \midrule
        No Profile 
        & 0.3275 & 0.4980 & 0.2500
        & 0.3505 & 0.5240 & 0.2250
        & 0.2950 & 0.4320 & 0.2665 \\
        RecAgent 
        & 0.2790 & 0.5080 & 0.2500
        & \textbf{0.7710} & \underline{0.8035} & 0.7165   
        & 0.4645 & 0.5670 & 0.5165 \\
        Agent4Rec 
        & \underline{0.4120} & 0.5605 & 0.3625
        & 0.6895 & 0.7465 & \underline{0.7700}
        & \textbf{0.6185} & \textbf{0.7000} & 0.5665 \\
        \midrule
        Semantic Merge 
        & 0.4020 & \underline{0.5800} & \underline{0.3835}
        & \underline{0.7705} & {0.7950} & \textbf{0.8000}
        & 0.5610 & 0.6740 & \underline{0.6000} \\
        APG4RecSim
        & \textbf{0.4300} & \textbf{0.5850} & \textbf{0.4165}
        & {0.7385} & \textbf{0.8105} & \textbf{0.8000}
        & \underline{0.5790} & \underline{0.6790} & \textbf{0.6335} \\
        \bottomrule
    \end{tabular}
\end{table*}

\begin{table}
    \centering
    \caption{Micro-level (RMSE) and macro-level (JSD) rating performance comparison across different profile-generation methods. We evaluate (1) an empty profile only, (2) raw recent interactions as context only, (3) profiles generated by RecAgent, (4) profiles generated by Agent4Rec, (5) APG4RecSim with semantic merging only, and (6) APG4RecSim with hybrid semantic–causal merging. The first two rows represent agents operating without generated profiles. Among the profile-based methods, the best performance is highlighted in bold, and the second-best is underlined.}
    \label{tab:rating_2}
    \begin{tabular}{c|cc|cc|cc}
        \toprule
        \multirow{2}{*}{} 
        & \multicolumn{2}{c}{ML-1M} & \multicolumn{2}{|c}{Book} & \multicolumn{2}{|c}{Beauty} \\
        & Micro$\downarrow$ & Macro$\downarrow$ & Micro$\downarrow$ & Macro$\downarrow$ & Micro$\downarrow$ & Macro$\downarrow$   \\
        \midrule
        1 
        & 1.0969 & 0.3241
        & 1.0145 & 0.2758
        & 1.2504 & 0.4376 \\
        2 
        & 1.0888 & 0.3157
        & 1.0166 & 0.2780
        & 1.2486 & 0.4368 \\
        \midrule
        3 
        & 1.1704 & 0.2863
        & 1.1181 & 0.2947
        & \textbf{1.2562} & \underline{0.4447} \\
        4 
        & 1.3068 & \textbf{0.2162}
        & 1.1498 & \textbf{0.1795}
        & 1.4290 & 0.4595 \\
        5 
        & \underline{1.1603} & 0.2515
        & \underline{1.1113} & 0.2276
        & 1.4475 & 0.4538 \\
        6
        & \textbf{1.1225} & \underline{0.2507}
        & \textbf{1.1028} & \underline{0.2113}
        & \underline{1.3967} & \textbf{0.4404} \\
        \bottomrule
    \end{tabular}
\end{table}

Table \ref{tab:discrimination}, \ref{tab:ranking}, and \ref{tab:rating_2} present the overall performance of different profile-generation methods across ranking, discrimination, and rating tasks on three benchmark datasets. In these tables, the best results are highlighted in bold, and the second-best results are underlined.
Note that Semantic Merge refers to the ablated version of our framework comprising only the first two stages (Initialisation and Semantic Consolidation). APG4RecSim denotes the complete proposed framework, incorporating the Causal Mapping and Refinement stage.

We observe that performance varies across metrics and tasks, and the proposed framework achieves the best performance on 16 out of 24 evaluation metrics and ranks second on 7 out of 24, demonstrating consistently strong performance across tasks and datasets.
APG4RecSim consistently delivers the strongest overall performance across tasks and datasets, achieving strong results across all tasks and datasets without sacrificing performance on any single evaluation dimension. In particular, our method maintains competitive or superior results across both micro- and macro-level evaluation metrics, demonstrating robust and well-balanced effectiveness rather than optimising for a single task or metric. This behaviour reflects the design goal of the framework, which prioritises task-aligned and generalisable profile quality over metric-specific optimisation. 
These results indicate that the proposed automated and context-aware profile generation framework provides a more reliable and transferable foundation for agent-based recommender system simulation than existing profile-generation baselines (\textbf{RQ1 and RQ2}).

\subsubsection{Discrimination Task}
Table \ref{tab:discrimination} summarises the performance of different profile-generation methods on the discrimination task across three benchmark datasets. Performance is evaluated using Overlap Ratio under varying levels of difficulty, requiring the agent to identify 1, 3, or 5 ground-truth items within a candidate list of 10 items (denoted as 1:10, 3:10, and 5:10 settings). 
The proposed approaches, Semantic Merge and APG4RecSim, generally demonstrate superior discrimination capabilities compared to the baseline methods across most dataset settings.
APG4RecSim achieves the best performance (bolded values) in 7 out of the 9 experimental scenarios. Semantic Merge frequently secures the second-best performance (underlined values), particularly in the tougher 1:10 settings across all datasets.

\subsubsection{Ranking Task}
Table \ref{tab:ranking} reports results of the ranking task on ML-1M, Amazon-Book, and Amazon-Beauty using nDCG@5, nDCG@10, and HR@3. Overall, APG4RecSim achieves the most consistent and robust performance across datasets, outperforming profile-free baselines and existing LLM-based profile generators.
On ML-1M, it delivers the best top-ranked relevance, as reflected by nDCG@5 and HR@3. On Amazon-Book, APG4RecSim attains the highest nDCG@10 and matches the best HR@3, demonstrating superior performance at deeper ranking positions. On Amazon-Beauty, it consistently achieves the best or near-best results across all metrics, confirming effective cross-domain generalisation. While not dominating every individual metric, APG4RecSim provides the most reliable and well-balanced ranking performance overall.

\subsubsection{Rating Task}
Table \ref{tab:rating_2} presents rating performance across ML-1M, Amazon-Book, and Amazon-Beauty, evaluated at both the micro level using RMSE and the macro level using JSD.
Except for the first two profile-free settings, LLM agents equipped with our generated profiles achieve the best overall rating performance, effectively balancing micro-level prediction accuracy with macro-level distribution consistency.
While some methods attain competitive RMSE, they often suffer from higher JSD, indicating misaligned global rating behaviour. In contrast, APG4RecSim, particularly with hybrid semantic–causal merging, consistently yields the lowest JSD across domains, demonstrating superior population-level rating alignment without sacrificing item-level accuracy.
The results reveal clear domain-dependent patterns, with different datasets exhibiting distinct trade-offs between micro-level accuracy and macro-level distribution alignment, underscoring the importance of adaptable profile generation rather than domain-specific heuristics.

We observe that agents conditioned on (1) an empty profile and (2) raw historical interactions achieve competitive performance on the micro-level RMSE metric, yet perform poorly on the macro-level distributional metric measured by JSD. Visualisation of the rating distributions on the Amazon-Books dataset, as shown in Figure \ref{fig:rating_distribution_book}, reveals that both settings produce ratings almost exclusively within the range of 3–5. 
This conservative rating behaviour yields low point-wise error for individual items which benefits RMSE but fails to reproduce the true rating distribution, resulting in substantial divergence at the population level.
This pattern suggests that, in the absence of structured and explicit user profiles, LLM-based agents tend to default to ``safe'' or neutral rating strategies that reflect general priors rather than personalised preference expression. While such behaviour is sufficient to minimise per-item prediction error, it masks the lack of calibration and diversity in simulated ratings, thereby undermining the realism of the simulation at the aggregate level. In contrast, profile-conditioned agents are better able to express both positive and negative preferences, leading to rating distributions that more closely match the ground truth despite occasionally incurring higher point-wise errors.

Overall, these results confirm that context-aware automated profiles lead to better behavioural alignment in rating simulation across domains, validating both the effectiveness and generalisability of our approach.

\subsection{Impact of Initialisation Examples}
\begin{figure}[t]
	\centering
    \begin{minipage}[c]{0.495\linewidth}
		\centering
        \includegraphics[width=\linewidth]{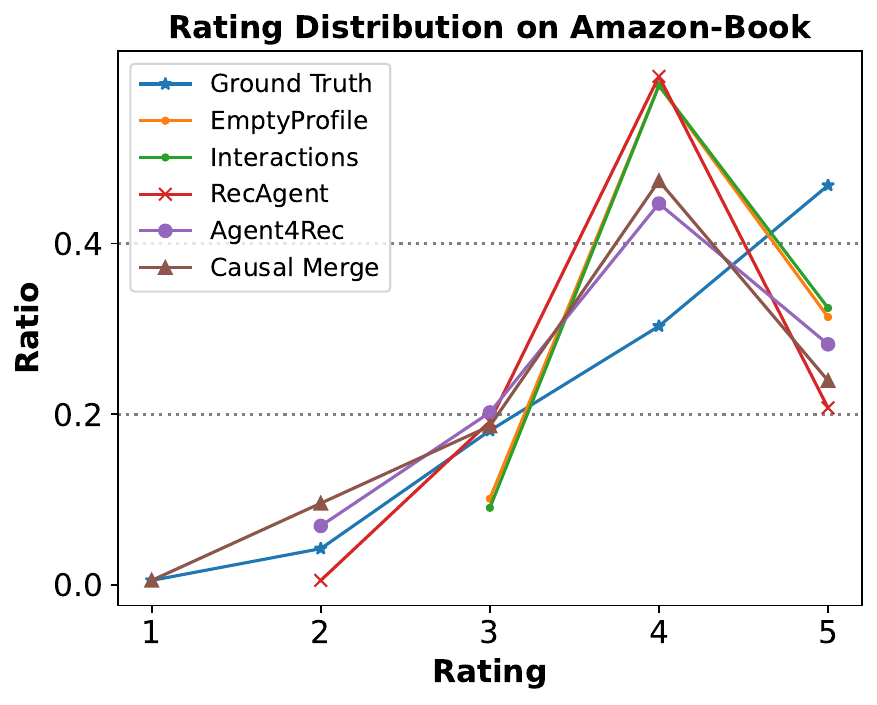}
		\subcaption{}
		\label{fig:rating_distribution_book}
	\end{minipage}
    \begin{minipage}[c]{0.495\linewidth}
		\centering
        \includegraphics[width=\linewidth]{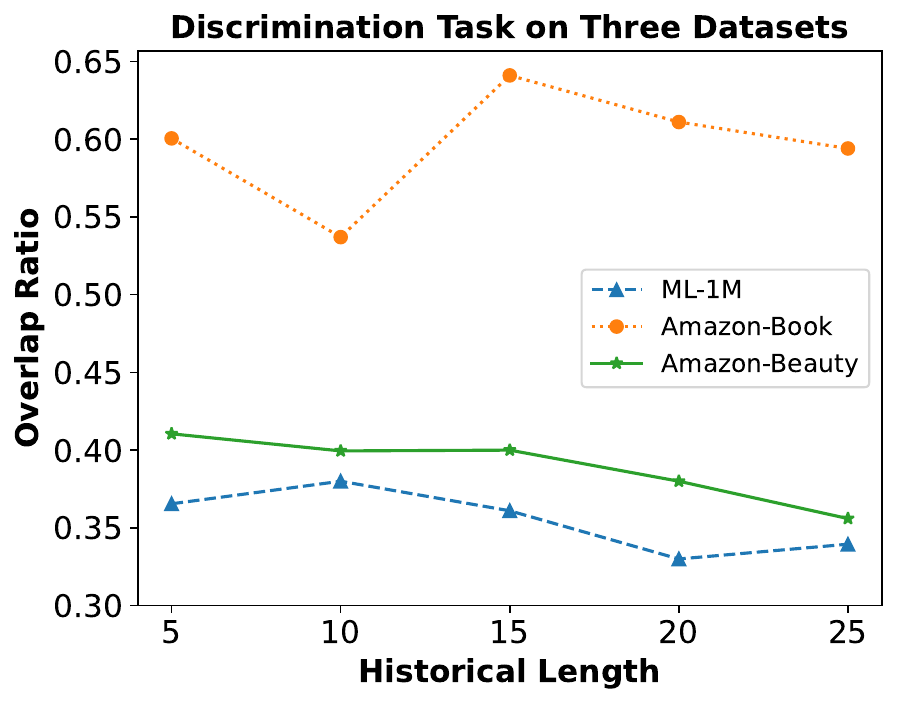}
		\subcaption{}
		\label{fig:interactions_on_3_datasets}
	\end{minipage}
    \begin{minipage}[c]{0.495\linewidth}
		\centering
		\includegraphics[width=\linewidth]{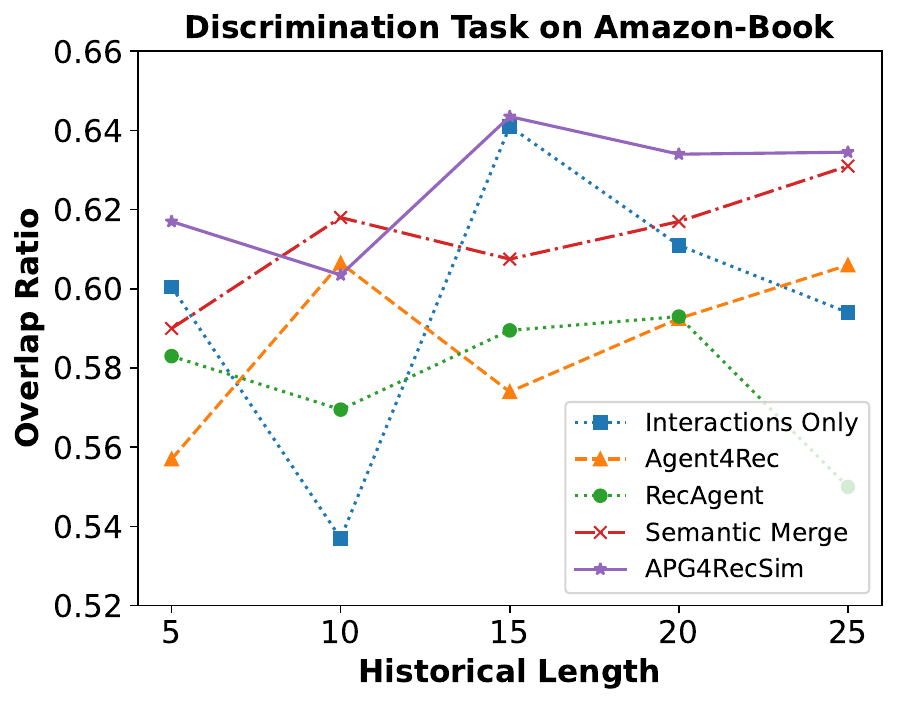}
		\subcaption{}
		\label{fig:interactions_on_book}
	\end{minipage}
    \begin{minipage}[c]{0.495\linewidth}
		\centering
		\includegraphics[width=\linewidth]{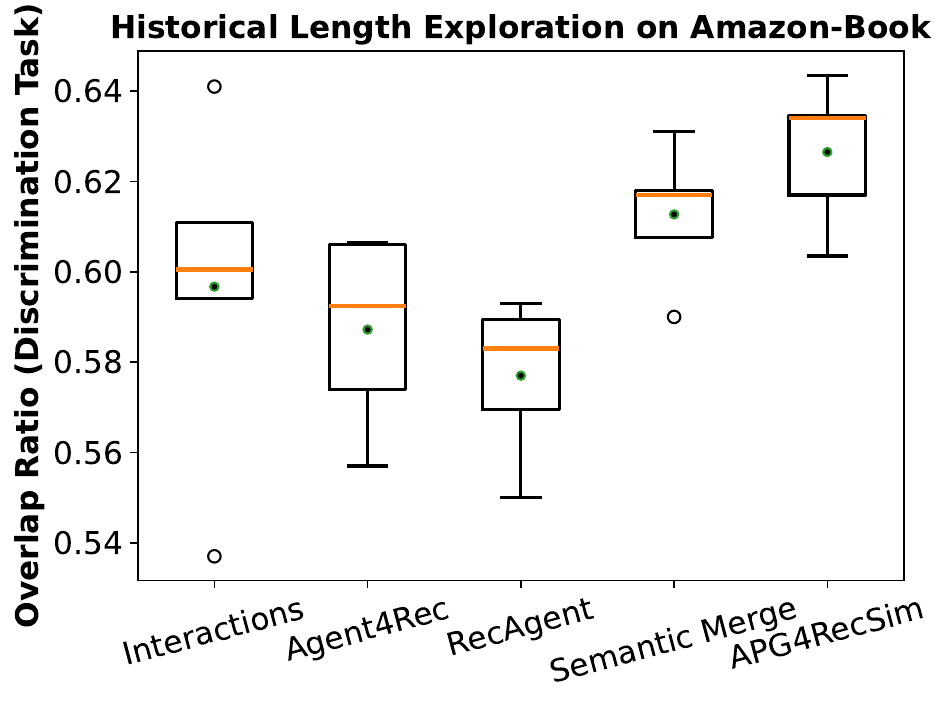}
		\subcaption{}
		\label{fig:interactions_on_book_boxplot}
	\end{minipage}
    \begin{minipage}[c]{0.495\linewidth}
		\centering
		\includegraphics[width=\linewidth]{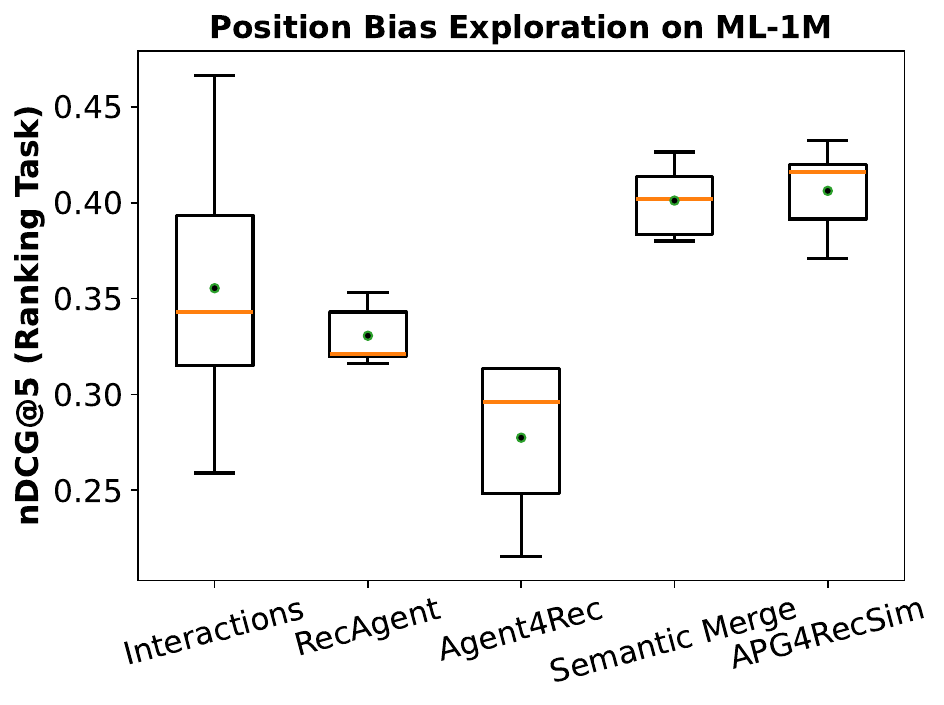}
		\subcaption{}
		\label{fig:position_bias}
	\end{minipage}
    \begin{minipage}[c]{0.495\linewidth}
		\centering
		\includegraphics[width=\linewidth]{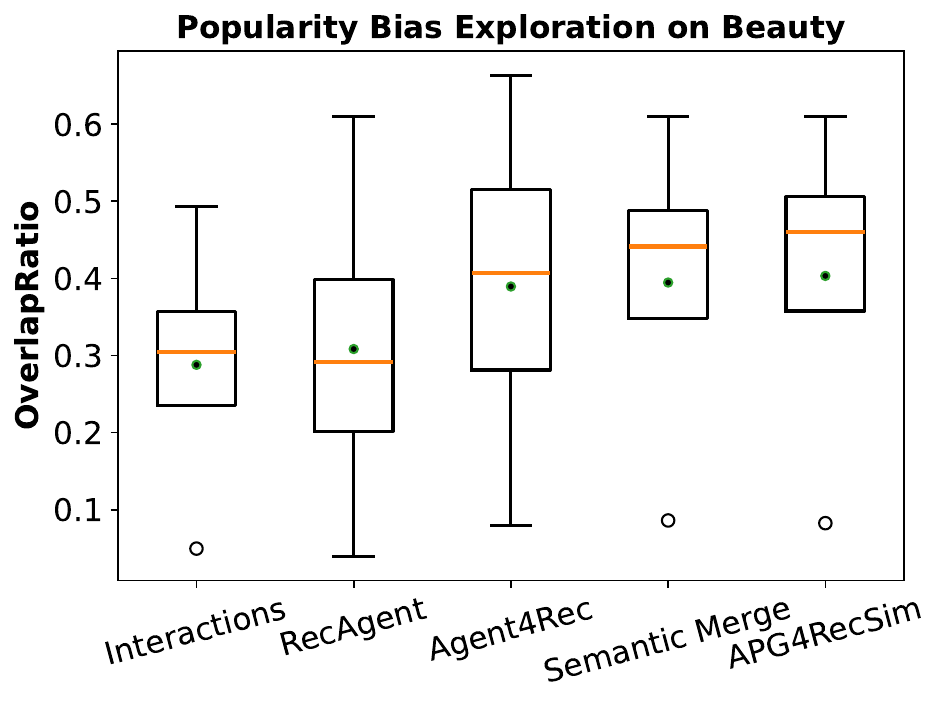}
		\subcaption{}
		\label{fig:popularity_bias}
	\end{minipage}
    \caption{
    (a) visualisation of the rating distribution on the Amazon-Book dataset;
    (b) impact of interaction history length on discriminative performance across multiple datasets; 
    (c, d) discriminative performance of generated profiles on the Amazon-Book dataset across varying interaction history lengths;
    (e) evaluation of robustness to position bias on the ML-1M dataset for a ranking task;
    (f) evaluation of robustness to popularity bias on the Amazon-Beauty dataset for a discrimination task (3 from 10).
    }
\end{figure}

We conducted a sensitivity analysis to evaluate the impact of input history length on the model's performance in the discrimination task across diverse recommender datasets. In this setup, the sequence of past user interactions serves as the contextual prompt, enabling the LLM to infer underlying preferences. 
By systematically varying the number of raw historical items provided in the context window, we analyse how the depth of interaction history correlates with the model's predictive accuracy.
Figure \ref{fig:interactions_on_3_datasets} illustrates the impact of input history length on discrimination accuracy across the ML-1M, Amazon-Book, and Beauty datasets.
We observe that discrimination performance drops as the historical length increases. 
Results from the Amazon-Book dataset identify a peak corresponding to a context length of 15 items, yielding a maximal overlap ratio of 0.641. The model appears more sensitive to a distinct lack of context than to an overabundance; reducing the interaction history from 15 to 10 results in a sharp drop of 10.4\% (to 0.537), whereas extending the history to 25 results in a more moderate decline of 4.7\% (to 0.594).
These findings highlight the fragility of direct-history prompting, where the optimal context length is highly sensitive and dataset-dependent.

Building on the insights regarding interaction history, we subsequently investigated the influence of the number of initialisation examples on profile generation task. Focusing on the Amazon-Book dataset, we systematically varied the count of provided examples to determine how the volume of few-shot demonstrations affects the fidelity and discriminative power of the generated user profiles.

\paragraph{Trend Analysis} 
The Interactions Only baseline (blue square) exhibits significant volatility. It suffers a sharp performance drop when history is short (10 items) and declines again as history extends (20–25 items), illustrating a susceptibility to both the ``cold-start'' problem and the ``lost-in-the-middle'' phenomenon, in which excessive context dilutes salient signals. In contrast, APG4RecSim (purple star) maintains a stable and superior trajectory. Notably, at the shortest history length (5 items), APG4RecSim achieves an overlap ratio of $\approx 0.62$, significantly outperforming the raw interaction baseline ($\approx 0.60$), confirming that our generated profiles effectively distil latent traits to provide a ``warm start'' even with sparse data.
\paragraph{Stability Analysis}
The box plot, as shown in Figure \ref{fig:interactions_on_book_boxplot}, aggregates the performance variance across all tested history lengths. APG4RecSim demonstrates superior robustness, achieving not only the highest median performance but also the tightest inter-quartile range. Conversely, the Interactions Only and RecAgent baselines show a much wider spread, indicating that their effectiveness is highly dependent on the specific length of the input history. This confirms that APG4RecSim's structured profiling decouples simulation quality from the variability of user history length, ensuring consistent behavioural alignment.

\begin{table*}
    \centering
    \caption{Generalisation results of the proposed automated profile generation framework on ML-1M and Amazon-Book across different LLMs. Discrimination ability is assessed using Overlap Ratio, and ranking quality is evaluated with nDCG@10 and Hit Rate@3. Best results are bolded. }
    \label{tab:different_llms_generalisation}
    \begin{tabular}{cc|cccc|cccc}
        \toprule
        \multirow{3}{*}{LLM} & \multirow{3}{*}{Model}
        & \multicolumn{4}{c}{ML-1M} & \multicolumn{4}{|c}{Book} \\
        &  & \multicolumn{2}{c}{Discrimination} & \multicolumn{2}{c|}{Ranking} & \multicolumn{2}{c}{Discrimination} & \multicolumn{2}{c}{Ranking} \\
        &  & 3:10 & 5:10 & nDCG@10 & HR@3 & 3:10 & 5:10 & nDCG@10 & HR@3 \\
        \midrule
        \multirow{4}{*}{Llama3.3-70B-Instruct}
        & RecAgent       
        & 0.2930 & 0.5000 & 0.5365 & 0.3000
        & 0.5935 & 0.7460 & 0.7160 & 0.6835 \\
        & Agent4Rec      
        & 0.3225 & 0.5440 & 0.5690 & 0.3500
        & 0.5665 & 0.7500 & 0.7365 & 0.7165 \\
        & Semantic Merge 
        & \textbf{0.3665} & 0.5540 & \textbf{0.5725} & 0.3335
        & 0.6170 & \textbf{0.7800} & 0.7480 & 0.7000 \\
        & APG4SimRec   
        & 0.3525 & \textbf{0.5760} & 0.5585 & \textbf{0.3510}
        & \textbf{0.6435} & 0.7500 & \textbf{0.7505} & \textbf{0.7500} \\
        \midrule
        \multirow{4}{*}{gpt-5.1-2025-11-13}
        & RecAgent
        & 0.3655 & \textbf{0.5440} & \textbf{0.4980} & 0.2835
        & 0.6035 & \textbf{0.7680} & 0.7455 & 0.7170 \\
        & Agent4Rec
        & 0.3155 & 0.4880 & 0.4750 & 0.2500
        & 0.6010 & 0.7600 & 0.7510 & 0.7000 \\
        & Semantic Merge 
        & 0.3365 & 0.4940 & \textbf{0.4980} & 0.2835
        & 0.6005 & 0.7200 & \textbf{0.8190} & \textbf{0.7500} \\
        & APG4SimRec  
        & \textbf{0.3665} & 0.5410 & 0.4870 & \textbf{0.3170}
        & \textbf{0.6275} & 0.7540 & 0.8080 & \textbf{0.7500} \\
        \midrule
        \multirow{4}{*}{DeepSeek-V3.2}
        & RecAgent  
        & 0.3035 & \textbf{0.5480} & 0.5110 & 0.2165 
        & 0.5695 & \textbf{0.7660} & 0.7120 & 0.5665 \\
        & Agent4Rec
        & 0.3265 & 0.5280 & 0.5235 & 0.2665 
        & \textbf{0.6315} & 0.7540 & 0.7200 & 0.6500 \\
        & Semantic Merge
        & 0.3330 & 0.5240 & 0.5090 & 0.3000
        & 0.6210 & 0.7580 & 0.7425 & 0.6165 \\
        & APG4SimRec
        & \textbf{0.3920} & 0.5380 & \textbf{0.5415} & \textbf{0.3165}
        & 0.6305 & 0.7440 & \textbf{0.7775} & \textbf{0.7335} \\
        \bottomrule
    \end{tabular}
\end{table*}

\subsection{Robustness and Generalisation Across LLMs}
To assess the generalisability of APG4SimRec, we benchmark the framework on the ML-1M and Amazon-Book datasets across diverse LLM architectures. 
We instantiate the framework using three distinct backbones, including Llama-3.3-70B-Instruct, GPT-5.1, and DeepSeek-V3.2, to assess profile performance across ranking and discrimination tasks.
Following our main experimental setup, we evaluate discrimination performance using a $P$-out-of-$C$ protocol ($P \in \{3, 5\}, C=10$), employing Overlap Ratio to measure the agent's ability to identify multiple ground-truth items within the constrained candidate set. 
Ranking quality is assessed under a leave-one-out setting using nDCG@10 and Hit Rate@3 (HR@3) to quantify the recovery of held-out user preferences.

Comparison results are shown in Table \ref{tab:different_llms_generalisation}.
Our framework demonstrates consistency across LLM architectures on different datasets. APG4RecSim achieves the best performance in 14 out of 24 total evaluation settings (spanning 3 LLMs, 2 datasets, and 4 metrics) and secures the second-best result in 7 additional cases. 
This robust performance demonstrates that our task-aware causal profiling strategy generalises effectively, operating independently of the specific capabilities of the underlying LLM backbone.
For instance, APG4SimRec achieves the highest HR@3 scores on all settings. This suggests that the proposed causal mapping strategy effectively grounds the agent's decision-making process, mitigating the dependency on the idiosyncratic reasoning patterns of any single LLM.
Furthermore, the performance gains are most pronounced with open-weight models (Llama-3.3-70B-Instruct and DeepSeek-V3.2), suggesting that our structured profiling effectively augments the capabilities of accessible models, bridging the gap between efficient architectures and proprietary giants.

This experimental design allows us to systematically compare different LLM backbones and examine the robustness and transferability of the proposed profile generation framework across datasets and model choices. 
(\textbf{RQ3})

\subsection{Robustness against Position Bias }
Position bias describes the tendency of users to disproportionately interact with top-ranked items regardless of their actual relevance or quality \cite{hofmann2014effects,collins2018position,biasRecSys2023}.
In LLM-agent simulations, this input-order sensitivity (or primacy effect) influences the preference evaluation, as high performance metrics may reflect mere position compliance rather than genuine interest \cite{DyTa4Rec}.

Figure \ref{fig:position_bias} evaluates the ranking performance (nDCG@5) across agents with different profiles generation methods on ML-1M dataset. 
The results indicate a significant performance gap between our proposed approach and the baselines.
The profiles generated by APG4RecSim achieve the highest overall performance, with a median nDCG@5 of approximately 0.415 and the highest mean score (indicated by the green dot). The Semantic Merge profiles follow closely as the second-best, with a median near 0.40. Both proposed methods substantially outperform all three baselines.
Furthermore, the box plot distributions show that while interactions only and Agent4Rec profile as the context exhibit high variance (large inter-quartile ranges and long whiskers), indicating inconsistent performance depending on the item's position, our APG4RecSim and semantic merging profiles maintain relatively compact distributions, suggesting stable decision-making regardless of item placement.

The inferior performance and high variance of the baselines indicate significant susceptibility to position bias, suggesting that agents relying on unstructured profiles or raw history struggle to decouple intrinsic item relevance from presentation order. Conversely, the stability of the agents with APG4RecSim profiles validates our structured methodology. By consolidating attributes and mapping them to explicit decision path nodes, these profiling methods ensure items are evaluated on semantic alignment rather than superficial ordering cues. The leading performance of the APG4RecSim method further indicates that explicit causal modelling within the user profile provides an effective mechanism for isolating true preferences from positional noise.

\subsection{Robustness against Popularity Bias}
Popularity bias is a pervasive issue in recommender systems, where algorithms disproportionately expose a small set of popular items while neglecting the majority of niche items in the ``long tail'' \cite{popularitybias2021,popularitybias2021Wei}. 
In the context of user behaviour simulation, popularity bias manifests as the simulated agent systematically preferring high-exposure items over personally relevant ones, regardless of the specific user profile being simulated.
Figure \ref{fig:popularity_bias} illustrates the performance of various profile generation methodologies on the Amazon-Beauty dataset. The y-axis represents the Overlap Ratio, where a higher ratio generally indicates that the method is less affected by popularity bias, as it maintains consistency across different sampling strategies. The x-axis displays different methodologies for profile generation. The box plots show the robustness of each method's Overlap Ratio when subjected to different negative item sampling strategies based on popularity.

The proposed methods, Semantic Merge and APG4RecSim, generally exhibit higher median Overlap Ratios compared to the baseline methods.
RecAgent profile and providing interactions only as the context have the lowest median Overlap Ratios, indicating that their performance is more significantly impacted by popularity bias. They also show a wider range of performance, suggesting lower robustness across different sampling strategies. Agent4Rec has a higher median than the other two baselines but exhibits a very large inter-quartile range, showing significant variability in its performance depending on the sampling strategy.

The results demonstrate that the proposed profile generation methods, Semantic Merge and APG4RecSim, are more robust to popularity bias in a discrimination task on the Amazon-Beauty dataset compared to baseline methods. They achieve higher median Overlap Ratios and demonstrate greater stability across different negative item sampling strategies.

\begin{table}
    \centering
    \caption{Discrimination performance on ML-1M using isolated item attributes to probe parametric priors. All settings utilise an empty profile, providing the agent only with specific item metadata (Title, Genre, Rating, Popularity). We compare two negative sampling strategies: \textbf{Random} (uniform sampling) and \textbf{De-bias} (popularity–rating matched sampling) to neutralise global popularity and quality biases.
    }
    \label{tab:LLMmemory}
    \begin{tabular}{ccccc|cc}
        \toprule
         Sampling & Title & Genre & Rating & Pop. & 3:10(r) & 3:6(r) \\
        \midrule
        \multirow{7}{*}{Random} & Yes & No & No & No & 0.2930 & 0.5390 \\
         & No & Yes & No & No & 0.2365 & 0.4955 \\
         & No & No & Yes & No & 0.3495 & 0.6340 \\
         & No & No & No & Yes & \textbf{0.4900} & \textbf{0.7400} \\
         & Yes& Yes& No & No  & 0.2595 & 0.5335 \\
         & No & No & Yes& Yes & 0.4225 & 0.6340 \\
         & Yes& Yes& Yes& Yes & 0.3105 & 0.6340 \\
        \midrule
        \multirow{7}{*}{De-bias} & Yes & No & No & No & 0.2710 & 0.5115 \\
         & No & Yes & No & No & 0.2150 & \textbf{0.5560} \\
         & No & No & Yes & No & 0.2490 & 0.5155 \\
         & No & No & No & Yes & \textbf{0.3990} & 0.5395 \\
         & Yes& Yes& No & No  & 0.2605 & 0.5385 \\
         & No & No & Yes& Yes & 0.2490 & 0.5320 \\
         & Yes& Yes& Yes& Yes & 0.2660 & 0.4925 \\
        \bottomrule
    \end{tabular}
\end{table}

\subsection{Impact of Parametric Knowledge and Global Priors}
LLMs can retain memory of widely used benchmark datasets like ML-1M, implicitly encoding item metadata and popularity distributions within their parametric knowledge \cite{carlini2022quantifying,2025_LLM_ML1M,wang2025generalisation}. 
This raises a critical validity concern for simulation: observed agent performance may stem from dataset-specific priors or global popularity cues, rather than genuine reasoning grounded in the provided user profile. 
To investigate the extent to which these internal priors drive discrimination performance, we conduct a controlled ablation study using two negative sampling strategies:
\textbf{(1) Random Sampling:} Negative items are drawn uniformly at random from the uninteracted set.
\textbf{(2) De-bias Sampling:} To neutralise popularity and quality biases, negative items are selected to statistically match the average popularity and rating of the ground-truth positive items.
We restrict the agent's input to isolated item attributes (e.g., Title only, Popularity only) to isolate the impact of specific cues. This design effectively probes the agent's reliance on global statistics versus semantic content.

Table \ref{tab:LLMmemory} reports the results on ML-1M. 
Under random sampling, providing Popularity alone yields dominant performance, consistently outperforming semantic attributes such as title and genre in both 3:10 and 3:6 settings. 
Providing rating information also leads to competitive results, whereas combining multiple attributes does not reliably improve performance beyond the strongest single cue, indicating diminishing returns once salient global statistics are available. These patterns suggest that, in the absence of explicit user profiles, the LLM-based agent tends to rely on coarse item-level statistics rather than integrating complementary semantic signals.
However, under de-bias sampling, performance collapses across all configurations and the popularity advantage vanishes. No single attribute setting consistently dominates.  
This indicates that a considerable portion of the performance observed under random sampling can be attributed to differences in global item statistics, rather than fine-grained semantic reasoning or personalised signals. 
This contrast confirms that standard LLM agents rely heavily on spurious global statistics.
While this analysis provides only indirect evidence regarding the role of LLM parametric memory, the results highlight the sensitivity of discrimination performance on ML-1M to item-level priors. 

Overall, these findings motivate the need for controlled profile evaluation and support the role of APG4RecSim in encouraging task-aligned, user-specific reasoning that moves beyond statistics-driven heuristics.

\section{Limitation and Future Work}
We acknowledge three limitations that outline critical directions for future research.

First, as observed in the diagnostic analysis on ML-1M, LLM parametric memory and global item-level cues (e.g., popularity or rating statistics) may partially influence agent behaviour, which could confound the assessment of genuine profile-driven reasoning. Future work should incorporate stronger controls, such as item anonymisation, synthetic datasets, or less publicly exposed domains, to further disentangle parametric priors from profile-conditioned behaviour.

Second, our framework focuses on profile generation and consolidation, intentionally isolating profile quality from agent architecture. While this design choice enables controlled evaluation, it does not explore the interaction between profiles and more sophisticated agent components, such as long-term memory, planning modules, or multi-step decision policies. Extending APG4RecSim to end-to-end agent systems remains an important direction for future work.

Third, the causal mapping and refinement step is derived from LLM-based counterfactual reasoning, which remains susceptible to inherent model stochasticity. Future work could explore hybrid approaches that integrate lightweight statistical or learning-based causal estimators to further stabilise profile refinement.

\section{Conclusion}
In this work, we proposed APG4RecSim, a training-free framework that automates profile generation by extracting task-aware latent traits without manual schemas (RQ1). Across multiple tasks and domains, our method achieves superior behavioural alignment, effectively balancing micro-level precision with macro-level fidelity (RQ2). Crucially, our controlled analysis confirms the framework's robustness, demonstrating that structured, task-aligned profiling mitigates the agent's reliance on global popularity priors and position bias (RQ3). 

Overall, this study provides a promising direction for automating user profile generation from existing recommendation datasets, and lays the groundwork for more generalised, interpretable, and reliable LLM agent-based user behaviour simulation for recommender systems.

\begin{acks}
This research was conducted by the ARC Centre of Excellence for Automated Decision-Making and Society (ADM+S, CE200100005) and funded by the Australian Government through the Australian Research Council.
This research was undertaken with the assistance of computing resources from RACE Hub (RMIT AWS Cloud Supercomputing Hub).
\end{acks}

\balance
\bibliographystyle{ACM-Reference-Format}
\bibliography{references}

@String{Computing = "Computing" }

@String{Computer = "{IEEE} Computer" }

@String{Academic = "Academic Press" }

@String{Springer = "Springer-Verlag" }

@inproceedings{Bhattacharjee2024,
author = {Bhattacharjee, Ananya and Zeng, Yuchen and Xu, Sarah Yi and Kulzhabayeva, Dana and Ma, Minyi and Kornfield, Rachel and Ahmed, Syed Ishtiaque and Mariakakis, Alex and Czerwinski, Mary P and Kuzminykh, Anastasia and Liut, Michael and Williams, Joseph Jay},
title = {Understanding the Role of Large Language Models in Personalizing and Scaffolding Strategies to Combat Academic Procrastination},
year = {2024},
isbn = {9798400703300},
publisher = {Association for Computing Machinery},
address = {New York, NY, USA},
url = {https://doi.org/10.1145/3613904.3642081},
doi = {10.1145/3613904.3642081},
booktitle = {Proceedings of the 2024 CHI Conference on Human Factors in Computing Systems},
articleno = {15},
numpages = {18},
location = {Honolulu, HI, USA},
series = {CHI '24}
}

@inproceedings{zhang2024guided,
    title = "Guided Profile Generation Improves Personalization with Large Language Models",
    author = "Zhang, Jiarui",
    editor = "Al-Onaizan, Yaser  and
      Bansal, Mohit  and
      Chen, Yun-Nung",
    booktitle = "Findings of the Association for Computational Linguistics: EMNLP 2024",
    month = nov,
    year = "2024",
    address = "Miami, Florida, USA",
    publisher = "Association for Computational Linguistics",
    url = "https://aclanthology.org/2024.findings-emnlp.231/",
    doi = "10.18653/v1/2024.findings-emnlp.231",
    pages = "4005--4016"
}

@article{zhang2025survey,
  title={A survey of large language model empowered agents for recommendation and search: Towards next-generation information retrieval},
  author={Zhang, Yu and Qiao, Shutong and Zhang, Jiaqi and Lin, Tzu-Heng and Gao, Chen and Li, Yong},
  journal={arXiv preprint arXiv:2503.05659},
  year={2025}
}

@article{hou2024bridging,
  title={Bridging language and items for retrieval and recommendation},
  author={Hou, Yupeng and Li, Jiacheng and He, Zhankui and Yan, An and Chen, Xiusi and McAuley, Julian},
  journal={arXiv preprint arXiv:2403.03952},
  year={2024}
}

@article{movielens_1m,
author = {Harper, F. Maxwell and Konstan, Joseph A.},
title = {The MovieLens Datasets: History and Context},
year = {2015},
issue_date = {January 2016},
publisher = {Association for Computing Machinery},
address = {New York, NY, USA},
volume = {5},
number = {4},
issn = {2160-6455},
url = {https://doi.org/10.1145/2827872},
doi = {10.1145/2827872},
journal = {ACM Trans. Interact. Intell. Syst.},
month = dec,
articleno = {19},
numpages = {19}
}

@article{wang2025recagent,
  title={User behavior simulation with large language model-based agents},
  author={Wang, Lei and Zhang, Jingsen and Yang, Hao and Chen, Zhi-Yuan and Tang, Jiakai and Zhang, Zeyu and Chen, Xu and Lin, Yankai and Sun, Hao and Song, Ruihua and others},
  journal={ACM Transactions on Information Systems},
  volume={43},
  number={2},
  pages={1--37},
  year={2025},
  publisher={ACM New York, NY}
}

@inproceedings{gao2022kuairec,
  title={KuaiRec: A fully-observed dataset and insights for evaluating recommender systems},
  author={Gao, Chongming and Li, Shijun and Lei, Wenqiang and Chen, Jiawei and Li, Biao and Jiang, Peng and He, Xiangnan and Mao, Jiaxin and Chua, Tat-Seng},
  booktitle={Proceedings of the 31st ACM International Conference on Information \& Knowledge Management},
  pages={540--550},
  year={2022}
}

@inproceedings{agent4rec,
author = {Zhang, An and Chen, Yuxin and Sheng, Leheng and Wang, Xiang and Chua, Tat-Seng},
title = {On Generative Agents in Recommendation},
year = {2024},
isbn = {9798400704314},
publisher = {Association for Computing Machinery},
address = {New York, NY, USA},
url = {https://doi.org/10.1145/3626772.3657844},
doi = {10.1145/3626772.3657844},
booktitle = {Proceedings of the 47th International ACM SIGIR Conference on Research and Development in Information Retrieval},
pages = {1807–1817},
numpages = {11},
location = {Washington DC, USA},
series = {SIGIR '24}
}

@inproceedings{ma2025pub,
  author       = {Chenglong Ma and
                  Ziqi Xu and
                  Yongli Ren and
                  Danula Hettiachchi and
                  Jeffrey Chan},
  title        = {{PUB:} An LLM-Enhanced Personality-Driven User Behaviour Simulator
                  for Recommender System Evaluation},
  booktitle    = {Proceedings of the 48th International {ACM} {SIGIR} Conference on
                  Research and Development in Information Retrieval, {SIGIR}},
  pages        = {2690--2694},
  year         = {2025}
}

@inproceedings{zhou2020s3,
  author       = {Kun Zhou and
                  Hui Wang and
                  Wayne Xin Zhao and
                  Yutao Zhu and
                  Sirui Wang and
                  Fuzheng Zhang and
                  Zhongyuan Wang and
                  Ji{-}Rong Wen},
  title        = {S3-Rec: Self-Supervised Learning for Sequential Recommendation with
                  Mutual Information Maximization},
  booktitle    = {{CIKM} '20: The 29th {ACM} International Conference on Information
                  and Knowledge Management},
  pages        = {1893--1902},
  year         = {2020}
}

@inproceedings{zhang2024agentcf,
  author       = {Junjie Zhang and
                  Yupeng Hou and
                  Ruobing Xie and
                  Wenqi Sun and
                  Julian J. McAuley and
                  Wayne Xin Zhao and
                  Leyu Lin and
                  Ji{-}Rong Wen},
  title        = {AgentCF: Collaborative Learning with Autonomous Language Agents for
                  Recommender Systems},
  booktitle    = {Proceedings of the {ACM} on Web Conference 2024, {WWW}},
  pages        = {3679--3689},
  year         = {2024}
}

@inproceedings{carlini2022quantifying,
  author       = {Nicholas Carlini and
                  Daphne Ippolito and
                  Matthew Jagielski and
                  Katherine Lee and
                  Florian Tram{\`{e}}r and
                  Chiyuan Zhang},
  title        = {Quantifying Memorization Across Neural Language Models},
  booktitle    = {The Eleventh International Conference on Learning Representations,
                  {ICLR} 2023, Kigali, Rwanda, May 1-5, 2023},
  publisher    = {OpenReview.net},
  year         = {2023},
  url          = {https://openreview.net/forum?id=TatRHT\_1cK},
  bibsource    = {dblp computer science bibliography, https://dblp.org}
}

@inproceedings{2025_LLM_ML1M,
author = {Di Palma, Dario and Merra, Felice Antonio and Sfilio, Maurizio and Anelli, Vito Walter and Narducci, Fedelucio and Di Noia, Tommaso},
title = {Do LLMs Memorize Recommendation Datasets? A Preliminary Study on MovieLens-1M},
year = {2025},
isbn = {9798400715921},
publisher = {Association for Computing Machinery},
address = {New York, NY, USA},
url = {https://doi.org/10.1145/3726302.3730178},
doi = {10.1145/3726302.3730178},
booktitle = {Proceedings of the 48th International ACM SIGIR Conference on Research and Development in Information Retrieval},
pages = {2582–2586},
numpages = {5},
location = {Padua, Italy},
series = {SIGIR '25}
}

@inproceedings{hou2022UniSRec,
author = {Hou, Yupeng and Mu, Shanlei and Zhao, Wayne Xin and Li, Yaliang and Ding, Bolin and Wen, Ji-Rong},
title = {Towards Universal Sequence Representation Learning for Recommender Systems},
year = {2022},
isbn = {9781450393850},
publisher = {Association for Computing Machinery},
address = {New York, NY, USA},
url = {https://doi.org/10.1145/3534678.3539381},
doi = {10.1145/3534678.3539381},
booktitle = {Proceedings of the 28th ACM SIGKDD Conference on Knowledge Discovery and Data Mining},
pages = {585–593},
numpages = {9},
location = {Washington DC, USA},
series = {KDD '22}
}

@inproceedings{DyTa4Rec,
author = {Wanyan, Xinye and Hettiachchi, Danula and Ma, Chenglong and Xu, Ziqi and Chan, Jeffrey},
title = {Temporal-Aware User Behaviour Simulation with Large Language Models for Recommender Systems},
year = {2025},
isbn = {9798400720406},
publisher = {Association for Computing Machinery},
address = {New York, NY, USA},
url = {https://doi.org/10.1145/3746252.3760878},
doi = {10.1145/3746252.3760878},
booktitle = {Proceedings of the 34th ACM International Conference on Information and Knowledge Management},
pages = {5335–5339},
numpages = {5},
location = {Seoul, Republic of Korea},
series = {CIKM '25}
}

@inproceedings{bougie2025simuser,
    title = "{S}im{USER}: Simulating User Behavior with Large Language Models for Recommender System Evaluation",
    author = "Bougie, Nicolas  and
      Watanabe, Narimawa",
    editor = "Rehm, Georg  and
      Li, Yunyao",
    booktitle = "Proceedings of the 63rd Annual Meeting of the Association for Computational Linguistics (Volume 6: Industry Track)",
    month = jul,
    year = "2025",
    address = "Vienna, Austria",
    publisher = "Association for Computational Linguistics",
    url = "https://aclanthology.org/2025.acl-industry.5/",
    doi = "10.18653/v1/2025.acl-industry.5",
    pages = "43--60",
    ISBN = "979-8-89176-288-6"
}

@ARTICLE{jsd1991,
  author={Lin, J.},
  journal={IEEE Transactions on Information Theory}, 
  title={Divergence measures based on the Shannon entropy}, 
  year={1991},
  volume={37},
  number={1},
  pages={145-151},
  doi={10.1109/18.61115}}

@article{ontological2004,
author = {Middleton, Stuart E. and Shadbolt, Nigel R. and De Roure, David C.},
title = {Ontological user profiling in recommender systems},
year = {2004},
issue_date = {January 2004},
publisher = {Association for Computing Machinery},
address = {New York, NY, USA},
volume = {22},
number = {1},
issn = {1046-8188},
url = {https://doi.org/10.1145/963770.963773},
doi = {10.1145/963770.963773},
journal = {ACM Trans. Inf. Syst.},
month = jan,
pages = {54–88},
numpages = {35}
}

@inproceedings{wang2025generalisation,
    title={Generalization v.s. Memorization: Tracing Language Models{\textquoteright} Capabilities Back to Pretraining Data},
    author={Xinyi Wang and Antonis Antoniades and Yanai Elazar and Alfonso Amayuelas and Alon Albalak and Kexun Zhang and William Yang Wang},
    booktitle={The Thirteenth International Conference on Learning Representations},
    year={2025},
    url={https://openreview.net/forum?id=IQxBDLmVpT}
}

@inproceedings{popularitybias2021Wei,
author = {Wei, Tianxin and Feng, Fuli and Chen, Jiawei and Wu, Ziwei and Yi, Jinfeng and He, Xiangnan},
title = {Model-Agnostic Counterfactual Reasoning for Eliminating Popularity Bias in Recommender System},
year = {2021},
isbn = {9781450383325},
publisher = {Association for Computing Machinery},
address = {New York, NY, USA},
url = {https://doi.org/10.1145/3447548.3467289},
doi = {10.1145/3447548.3467289},
booktitle = {Proceedings of the 27th ACM SIGKDD Conference on Knowledge Discovery \& Data Mining},
pages = {1791–1800},
numpages = {10},
location = {Virtual Event, Singapore},
series = {KDD '21}
}

@inproceedings{popularitybias2021,
author = {Abdollahpouri, Himan and Mansoury, Masoud and Burke, Robin and Mobasher, Bamshad and Malthouse, Edward},
title = {User-centered Evaluation of Popularity Bias in Recommender Systems},
year = {2021},
isbn = {9781450383660},
publisher = {Association for Computing Machinery},
address = {New York, NY, USA},
url = {https://doi.org/10.1145/3450613.3456821},
doi = {10.1145/3450613.3456821},
booktitle = {Proceedings of the 29th ACM Conference on User Modeling, Adaptation and Personalization},
pages = {119–129},
numpages = {11},
location = {Utrecht, Netherlands},
series = {UMAP '21}
}

@inproceedings{chakraborty2017makes,
  title={Who makes trends? understanding demographic biases in crowdsourced recommendations},
  author={Chakraborty, Abhijnan and Messias, Johnnatan and Benevenuto, Fabricio and Ghosh, Saptarshi and Ganguly, Niloy and Gummadi, Krishna},
  booktitle={Proceedings of the International AAAI Conference on Web and Social Media},
  volume={11},
  number={1},
  pages={22--31},
  year={2017}
}

@inproceedings{zhang2020evaluating,
  title={Evaluating conversational recommender systems via user simulation},
  author={Zhang, Shuo and Balog, Krisztian},
  booktitle={Proceedings of the 26th acm sigkdd international conference on knowledge discovery \& data mining},
  pages={1512--1520},
  year={2020}
}

@inproceedings{evalrec25xu,
author = {Xu, Ziqi and Ma, Chenglong and Ren, Yongli and Chan, Jeffrey and Shao, Wei and Xia, Feng},
title = {Towards Better Evaluation of Recommendation Algorithms with Bi-directional Item Response Theory},
year = {2025},
isbn = {9798400713316},
publisher = {Association for Computing Machinery},
address = {New York, NY, USA},
url = {https://doi.org/10.1145/3701716.3715540},
doi = {10.1145/3701716.3715540},
booktitle = {Companion Proceedings of the ACM on Web Conference 2025},
pages = {1455–1459},
numpages = {5},
location = {Sydney NSW, Australia},
series = {WWW '25}
}

@inproceedings{beyond2025jin,
    title = "Beyond Static Testbeds: An Interaction-Centric Agent Simulation Platform for Dynamic Recommender Systems",
    author = "Jin, Song  and
      Zhang, Juntian  and
      Liu, Yuhan  and
      Zhang, Xun  and
      Zhang, Yufei  and
      Yin, Guojun  and
      Jiang, Fei  and
      Lin, Wei  and
      Yan, Rui",
    editor = "Christodoulopoulos, Christos  and
      Chakraborty, Tanmoy  and
      Rose, Carolyn  and
      Peng, Violet",
    booktitle = "Proceedings of the 2025 Conference on Empirical Methods in Natural Language Processing",
    month = nov,
    year = "2025",
    address = "Suzhou, China",
    publisher = "Association for Computational Linguistics",
    url = "https://aclanthology.org/2025.emnlp-main.956/",
    doi = "10.18653/v1/2025.emnlp-main.956",
    pages = "18903--18920",
    ISBN = "979-8-89176-332-6"
}

@inproceedings{RecUserSim2025Chen,
author = {Chen, Luyu and Dai, Quanyu and Zhang, Zeyu and Feng, Xueyang and Zhang, Mingyu and Tang, Pengcheng and Chen, Xu and Zhu, Yue and Dong, Zhenhua},
title = {RecUserSim: A Realistic and Diverse User Simulator for Evaluating Conversational Recommender Systems},
year = {2025},
isbn = {9798400713316},
publisher = {Association for Computing Machinery},
address = {New York, NY, USA},
url = {https://doi.org/10.1145/3701716.3715258},
doi = {10.1145/3701716.3715258},
booktitle = {Companion Proceedings of the ACM on Web Conference 2025},
pages = {133–142},
numpages = {10},
location = {Sydney NSW, Australia},
series = {WWW '25}
}

@article{llmmemory2025Zhang,
author = {Zhang, Zeyu and Dai, Quanyu and Bo, Xiaohe and Ma, Chen and Li, Rui and Chen, Xu and Zhu, Jieming and Dong, Zhenhua and Wen, Ji-Rong},
title = {A Survey on the Memory Mechanism of Large Language Model-based Agents},
year = {2025},
issue_date = {November 2025},
publisher = {Association for Computing Machinery},
address = {New York, NY, USA},
volume = {43},
number = {6},
issn = {1046-8188},
url = {https://doi.org/10.1145/3748302},
doi = {10.1145/3748302},
journal = {ACM Trans. Inf. Syst.},
month = sep,
articleno = {155},
numpages = {47}
}

@INPROCEEDINGS{Profiling2025Wang,
  author={Wang, Hanpeng and Yang, Zijiang},
  booktitle={2025 International Conference on Control, Automation and Diagnosis (ICCAD)}, 
  title={A Multi-Agent Approach to Investor Profiling Using Large Language Models}, 
  year={2025},
  volume={},
  number={},
  pages={1-6},
  doi={10.1109/ICCAD64771.2025.11099326}}

@article{hu2024quantifying,
  title={Quantifying the persona effect in llm simulations},
  author={Hu, Tiancheng and Collier, Nigel},
  journal={arXiv preprint arXiv:2402.10811},
  year={2024}
}

@article{zhao2024recommender,
  title={Recommender systems in the era of large language models (llms)},
  author={Zhao, Zihuai and Fan, Wenqi and Li, Jiatong and Liu, Yunqing and Mei, Xiaowei and Wang, Yiqi and Wen, Zhen and Wang, Fei and Zhao, Xiangyu and Tang, Jiliang and others},
  journal={IEEE Transactions on Knowledge and Data Engineering},
  volume={36},
  number={11},
  pages={6889--6907},
  year={2024},
  publisher={IEEE}
}

@inproceedings{deldjoo2024review,
  title={A review of modern recommender systems using generative models (gen-recsys)},
  author={Deldjoo, Yashar and He, Zhankui and McAuley, Julian and Korikov, Anton and Sanner, Scott and Ramisa, Arnau and Vidal, Ren{\'e} and Sathiamoorthy, Maheswaran and Kasirzadeh, Atoosa and Milano, Silvia},
  booktitle={Proceedings of the 30th ACM SIGKDD conference on Knowledge Discovery and Data Mining},
  pages={6448--6458},
  year={2024}
}

@inproceedings{hofmann2014effects,
  title={Effects of position bias on click-based recommender evaluation},
  author={Hofmann, Katja and Schuth, Anne and Bellogin, Alejandro and De Rijke, Maarten},
  booktitle={European Conference on Information Retrieval},
  pages={624--630},
  year={2014},
  organization={Springer}
}

@inproceedings{collins2018position,
  title={Position bias in recommender systems for digital libraries},
  author={Collins, Andrew and Tkaczyk, Dominika and Aizawa, Akiko and Beel, Joeran},
  booktitle={International Conference on Information},
  pages={335--344},
  year={2018},
  organization={Springer}
}

@article{biasRecSys2023,
author = {Chen, Jiawei and Dong, Hande and Wang, Xiang and Feng, Fuli and Wang, Meng and He, Xiangnan},
title = {Bias and Debias in Recommender System: A Survey and Future Directions},
year = {2023},
issue_date = {July 2023},
publisher = {Association for Computing Machinery},
address = {New York, NY, USA},
volume = {41},
number = {3},
issn = {1046-8188},
url = {https://doi.org/10.1145/3564284},
doi = {10.1145/3564284},
journal = {ACM Trans. Inf. Syst.},
month = feb,
articleno = {67},
numpages = {39}
}

@inproceedings{genderbias2025,
author = {Krause, Thorsten and G\"{o}ritz, Lorena and Gratz, Robin},
title = {The Effect of Gender De-biased Recommendations — A User Study on Gender-specific Preferences},
year = {2025},
isbn = {9798400713941},
publisher = {Association for Computing Machinery},
address = {New York, NY, USA},
url = {https://doi.org/10.1145/3706598.3713155},
doi = {10.1145/3706598.3713155},
booktitle = {Proceedings of the 2025 CHI Conference on Human Factors in Computing Systems},
articleno = {1000},
numpages = {16},
location = {
},
series = {CHI '25}
}

@inproceedings{sigir2025Zhang,
author = {Zhang, Erhan and Wang, Xingzhu and Gong, Peiyuan and Yang, Zixuan and Mao, Jiaxin},
title = {Exploring Human-Like Thinking in Search Simulations with Large Language Models},
year = {2025},
isbn = {9798400715921},
publisher = {Association for Computing Machinery},
address = {New York, NY, USA},
url = {https://doi.org/10.1145/3726302.3730193},
doi = {10.1145/3726302.3730193},
booktitle = {Proceedings of the 48th International ACM SIGIR Conference on Research and Development in Information Retrieval},
pages = {2669–2673},
numpages = {5},
location = {Padua, Italy},
series = {SIGIR '25}
}

@inproceedings{sigir2025Ungruh,
author = {Ungruh, Robin and Bellog\'{\i}n, Alejandro and Pera, Maria Soledad},
title = {From Monolith to Mosaic: Uncovering Behavioral Differences for Choice Models in Recommender Systems Simulations},
year = {2025},
isbn = {9798400715921},
publisher = {Association for Computing Machinery},
address = {New York, NY, USA},
url = {https://doi.org/10.1145/3726302.3730199},
doi = {10.1145/3726302.3730199},
booktitle = {Proceedings of the 48th International ACM SIGIR Conference on Research and Development in Information Retrieval},
pages = {2717–2722},
numpages = {6},
location = {Padua, Italy},
series = {SIGIR '25}
}

@inproceedings{sigir2025Cai,
  author={Shihao Cai and Jizhi Zhang and Keqin Bao and Chongming Gao and Qifan Wang and Fuli Feng and Xiangnan He},
  title={Agentic Feedback Loop Modeling Improves Recommendation and User Simulation},
  year={2025},
  cdate={1735689600000},
  pages={2235-2244},
  url={https://doi.org/10.1145/3726302.3729893},
  booktitle={SIGIR}
}

@article{demographic2016recsys,
author = {Al-Shamri, Mohammad Yahya H.},
title = {User profiling approaches for demographic recommender systems},
year = {2016},
issue_date = {May 2016},
publisher = {Elsevier Science Publishers B. V.},
address = {NLD},
volume = {100},
number = {C},
issn = {0950-7051},
url = {https://doi.org/10.1016/j.knosys.2016.03.006},
doi = {10.1016/j.knosys.2016.03.006},
journal = {Know.-Based Syst.},
month = may,
pages = {175–187},
numpages = {13}
}

@inproceedings{lettingo2025Wang,
author = {Wang, Lu and Zhang, Di and Yang, Fangkai and Zhao, Pu and Liu, Jianfeng and Zhan, Yuefeng and Sun, Hao and Lin, Qingwei and Deng, Weiwei and Zhang, Dongmei and Sun, Feng and Zhang, Qi},
title = {LettinGo: Explore User Profile Generation for Recommendation System},
year = {2025},
isbn = {9798400714542},
publisher = {Association for Computing Machinery},
address = {New York, NY, USA},
url = {https://doi.org/10.1145/3711896.3737024},
doi = {10.1145/3711896.3737024},
booktitle = {Proceedings of the 31st ACM SIGKDD Conference on Knowledge Discovery and Data Mining V.2},
pages = {2985–2995},
numpages = {11},
location = {Toronto ON, Canada},
series = {KDD '25}
}

@inproceedings{Zhang_2025_LLMpowerSim4Rec,
author = {Zhang, Zijian and Liu, Shuchang and Liu, Ziru and Zhong, Rui and Cai, Qingpeng and Zhao, Xiangyu and Zhang, Chunxu and Liu, Qidong and Jiang, Peng},
title = {LLM-powered user simulator for recommender system},
year = {2025},
isbn = {978-1-57735-897-8},
publisher = {AAAI Press},
url = {https://doi.org/10.1609/aaai.v39i12.33456},
doi = {10.1609/aaai.v39i12.33456},
booktitle = {Proceedings of the Thirty-Ninth AAAI Conference on Artificial Intelligence and Thirty-Seventh Conference on Innovative Applications of Artificial Intelligence and Fifteenth Symposium on Educational Advances in Artificial Intelligence},
articleno = {1483},
numpages = {9},
series = {AAAI'25/IAAI'25/EAAI'25}
}

@inproceedings{chen2019generative,
  title={Generative adversarial user model for reinforcement learning based recommendation system},
  author={Chen, Xinshi and Li, Shuang and Li, Hui and Jiang, Shaohua and Qi, Yuan and Song, Le},
  booktitle={International conference on machine learning},
  pages={1052--1061},
  year={2019},
  organization={PMLR}
}

@inproceedings{UserIPTuning2025,
author = {Lu, Yusheng and Du, Zhaocheng and Li, Xiangyang and Jia, Pengyue and Wang, Yejing and Liu, Weiwen and Wang, Yichao and Guo, Huifeng and Tang, Ruiming and Dong, Zhenhua and Duan, Yongrui and Zhao, Xiangyu},
title = {Prompt Tuning as User Inherent Profile Inference Machine},
year = {2025},
isbn = {9798400720406},
publisher = {Association for Computing Machinery},
address = {New York, NY, USA},
url = {https://doi.org/10.1145/3746252.3761574},
doi = {10.1145/3746252.3761574},
booktitle = {Proceedings of the 34th ACM International Conference on Information and Knowledge Management},
pages = {5898–5906},
numpages = {9},
location = {Seoul, Republic of Korea},
series = {CIKM '25}
}

@inproceedings{Park2023Simulacra,
author = {Park, Joon Sung and O'Brien, Joseph and Cai, Carrie Jun and Morris, Meredith Ringel and Liang, Percy and Bernstein, Michael S.},
title = {Generative Agents: Interactive Simulacra of Human Behavior},
year = {2023},
isbn = {9798400701320},
publisher = {Association for Computing Machinery},
address = {New York, NY, USA},
url = {https://doi.org/10.1145/3586183.3606763},
doi = {10.1145/3586183.3606763},
booktitle = {Proceedings of the 36th Annual ACM Symposium on User Interface Software and Technology},
articleno = {2},
numpages = {22},
location = {San Francisco, CA, USA},
series = {UIST '23}
}

@article{radford2019language,
  title={Language models are unsupervised multitask learners},
  author={Radford, Alec and Wu, Jeffrey and Child, Rewon and Luan, David and Amodei, Dario and Sutskever, Ilya and others},
  journal={OpenAI blog},
  volume={1},
  number={8},
  pages={9},
  year={2019}
}

@STRING{academic= "Academic Press" }

@STRING{computer= "{IEEE} Computer" }

@STRING{computing="Computing" }

@STRING{springer= "Springer-Verlag" }




\end{document}